# 5G-ENABLED SMART MANUFACTURING

## A booklet by 5G-SMART

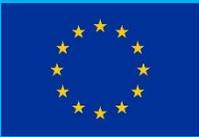

The 5G-SMART project has received funding from the European Union's Horizon 2020 research and innovation programme under grant agreement No 857008.

In this booklet the most important learnings and key results of 5G-SMART in the area of smart manufacturing are summarized.

Authors: Leefke Grosjean (Ericsson), Krister Landernäs (ABB), Berna Sayrac (Orange), Ognjen Dobrijevic (ABB), Niels König (Fraunhofer IPT), Davit Harutyunyan (Bosch), Dhruvin Patel (Ericsson), Jose F. Monserrat (UPV), Joachim Sachs (Ericsson)

Date: August 2022

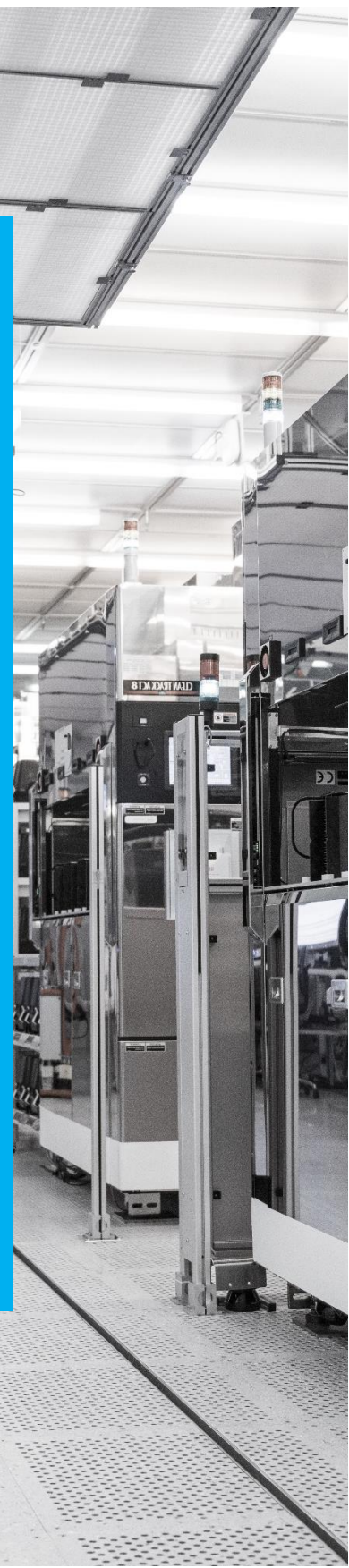

# Introduction

Today the smart manufacturing sector is undergoing a digital transformation addressing the challenges of reducing manual processes and increasing efficiency. 5G is as key enabler for the future manufacturing ecosystem termed Industry 4.0. Future factories will be characterized by flexible, modular production systems, requiring reliable and versatile communication and computations solutions. The journey towards this vision has already started but has gained even more attraction during the Covid-19 pandemic, which has shown the clear need for even greater flexibility and rapid adoption to changes in the manufacturing sector.

The fusion of a high-performance wireless communication infrastructure of 5G with the high degree of automation already present in today's factories is expected to unleash novel opportunities and lift the manufacturing industry to the next level. However, introducing 5G on the factory floor is not only about replacing cables; it is also about redefining and reinventing the wireless manufacturing ecosystem. 5G networked environments have the potential to create positive impact on a wide range of areas from worker safety to productivity enhancements, sustainability and more. Forward-looking enterprises already now investigate how the new capabilities coming with 5G-integrations open up new opportunities in automation. Condition monitoring or fleet management are only examples of areas that will boost productivity and efficiency when empowered by 5G.

The disruptive changes are however not going to happen overnight. 5G-enhanced smart manufacturing requires research, investigations, validations and evaluations. The project 5G-SMART has made a significant contribution to this challenge, building on the efforts in the area that have been made over the last years. 5G-enhanced smart manufacturing is now put into practice. With the validation trials in the different testbeds built up by the project, 5G-SMART has shown the feasibility of Industry 4.0 use cases empowered by 5G today, while at the same time contributing to the community with learnings and findings.

A critical element of success when transforming towards Industry 4.0 is to bring the right expert partners of the ecosystem together. This has been the mindset and starting point of the 5G-SMART project.

With its unique composition of the consortium, the project has managed to cover aspects from 5G

> **THE 5G-SMART PROJECT**
>
> 5G-SMART is an EU project that during a 3-years period from 2019-2022 has worked towards the goal of showing how 5G can boost smart manufacturing. The project was managed by Ericsson and ABB. The multidisciplinary team of 5G-SMART consisted of Information and Communications Technology (ICT) and 5G suppliers (Ericsson, Cumucore, T-systems Hungary), network operators (Orange), providers of wireless communication technologies and components (u-blox), operational technologies' suppliers (ABB, Bosch, Fraunhofer IPT, Marposs), factory operators (Bosch) and academia (Lund University, University of Valencia, Budapest University of Technology and Economics). The consortium has furthermore been closely interacting with 5G-ACIA. All three testbeds of 5G-SMART have been endorsed by 5G-ACIA (5G Alliance for Connected Industries and Automation).

integration into the ecosystem (use cases, business models, business relationships, deployment options) to actual validation, evaluation and demonstration of 5G-enhanced use cases. On top of that, the project paints the journey towards future optimization of 5G for smart manufacturing.

Figure 1 gives an overview of 5G-SMART, illustrating the topics addressed in the project.

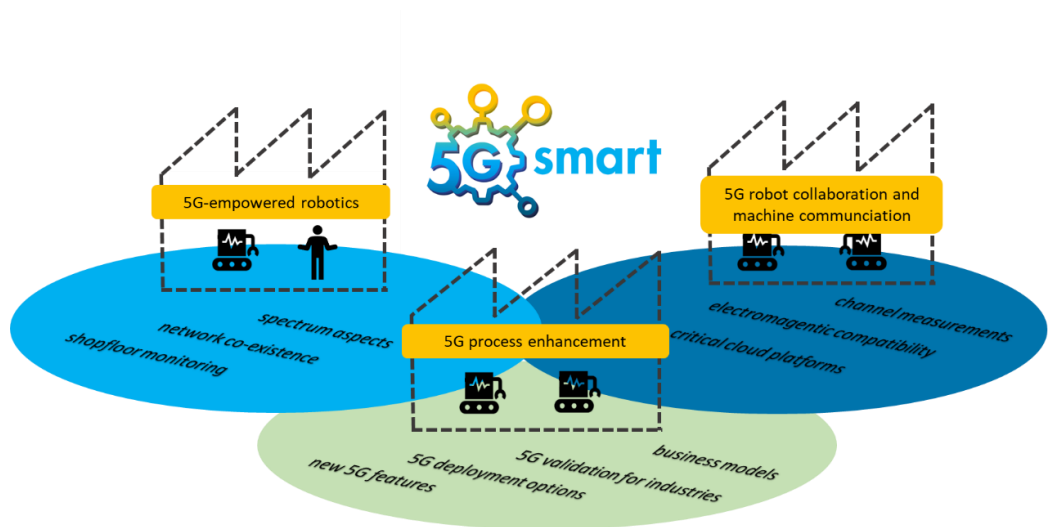

*Figure 1: 5G-SMART overview*

This booklet does by no means claim to give a complete overview of the topic but it illustrates and exemplifies key topics to consider in the transformation towards Industry 4.0 using the 5G-SMART project results, key learnings and findings. The terminology used for definitions and terms in this document can be found at [5GS20-CT]. Enjoy the reading!

Disclaimer
This work has been performed in the framework of the H2020 project 5G-SMART co-funded by the EU. This information reflects the consortium's view, but the consortium is not liable for any use that may be made of any of the information contained therein.

# Manufacturing use cases

The application areas of smart manufacturing contain a plethora of use cases with different requirements on the communication and computation infrastructure, which need to be fulfilled by the underlying 5G infrastructure. 5G-SMART has made an important contribution to better understanding the manufacturing sector's needs concerning 5G with the exploration and analysis of both trialed use cases and additional forward-looking smart manufacturing use cases. For all of the use cases, the benefits of using 5G and requirements and challenges have been assessed and described. The complete analysis is published in [5GS20-D110]. Table 1 lists all use cases that have been investigated in 5G-SMART together with their classification according to 3GPP [TR22.804]. The use cases 1-7 have been trialed in the 5G-SMART project and will be further explained later in the document.

| | 5G-SMART Use cases | Factory automation | Process automation | HMIs and Production IT | Logistics and warehousing | Monitoring and maintenance |
|---|---|---|---|---|---|---|
| 1 | 5G-Connected Robot and Remotely Supported Collaboration | X | | | X | |
| 2 | Machine Vision Assisted Real-time Human-Robot Interaction over 5G | X | | X | X | |
| 3 | 5G-Aided Visualization of the Factory Floor | X | | X | | X |
| 4 | 5G for Wireless Acoustic Workpiece Monitoring | X | X | | | X |
| 5 | 5G Versatile Multi-Sensor Platform for Digital Twin | X | X | | | X |
| 6 | Cloud-based Mobile Robotics | X | | | X | |
| 7 | Time Sensitive Networking (TSN)/Industrial LAN over 5G | X | | | | |
| 8 | 5G-Enabled Remote Expert | X | X | X | X | X |
| 9 | 5G Empowered Cross-domain and Inter-company Collaboration | X | X | X | X | X |
| 10 | Automated Guided Vehicle (AGV) Realtime Trajectory Adaption with AI for Smart Factories | X | X | | X | |
| 11 | 5G Enabled Metrology and Process Control across Machine and Factory Boundaries | X | X | | | X |
| 12 | 5G Enabled Seamless Device Plug and Play | X | X | X | | X |
| 13 | AI-assisted Production Quality Management | X | X | | | X |

*Table 1: Smart manufacturing use cases considered in 5G-SMART*

The key takeaways from the use cases analysis are summarized below.

> **KEY RESULTS AND FINDINGS**
> 
> - The analysis of the requirements and KPIs of the use cases clearly shows the need for a reliable, low-latency, high-performance wireless communication infrastructure in future factories. This holds for operations within a single factory or manufacturing site, which the trialed use cases focus on, and for more extended deployments, going beyond a single location or company, where the local 5G network has to inter-work with traditional public cellular networks.
> - Remote operations and monitoring have been identified as a key application of several use cases in smart manufacturing, demanding reliable wireless technology for their implementation. Remote operation is expected to becomean essential driver of 5G in the future, which already became apparent during the outbreak of Covid-19.
> - 5G technology will play an important role in realizing future Artificial Intelligence (AI) solutions for factory automation. Here, the bounded latency and scalability of 5G will enable easy and reliable connectivity to a large number of sensors in the factory. Edge cloud deployments will enable computational offloading allowing fast execution of the AI engines locally.

The methodology for validation and evaluation of the first seven use cases from Table 1, implemented in the testbeds built up by 5G-SMART, is described in the following section.

### METHODOLOGY FOR VALIDATION IN TRIALS

The methodology applied for validating, evaluating and demonstrating 5G capabilities in the industry field trials is outlined in Figure 2. All trials have followed the same approach, starting with a thorough analysis of the use cases, including their operational, functional, and performance requirements. Apart from the requirements, a key ambition for all testbeds has also been to understand the industry goals and economic benefits of enhancing the particular use cases with 5G. The use case characterization, floorplans and their characteristics have been the input for performing radio planning to determine a suitable 5G deployment to support the testbeds. Use case development and implementation have been the central part of the work. For each testbed, important Key Performance Indicators (KPIs) have been identified for validation and evaluation, and validation scenarios have been designed in order to quantify results. The results of the validation and evaluation have been documented. Use case executions have been demonstrated at various events.

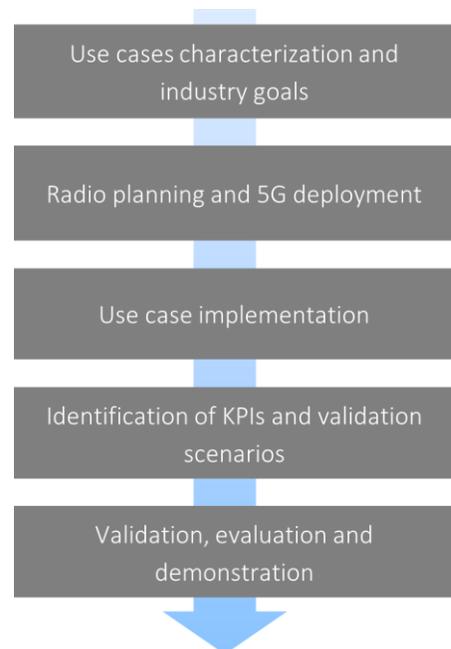

*Figure 2: Methodology for validation and evaluation*

# 5G integration into the ecosystem

### BUSINESS VALUE ASSESSMENT FRAMEWORK

Between 2021 and 2025, Industry 4.0 technologies are expected to increase the production industry's gross margins significantly. Applications such as closed-loop control, predictive maintenance, digital twins, augmented reality, and automated guided vehicles are expected to be essential enablers. A reliable, low-latency, high-performance wireless communication infrastructure plays a decisive role in these applications. While many production companies endorse the predictions of massive benefits from 5G for productivity, uncertainty is still inhibiting investments so far. The main reason is that a framework is missing to quantify potential improvements and allow an economic evaluation of 5G-enabled benefits for specific use cases from an end-user perspective.

5G-SMART has developed an assessment framework to quantify the business value of 5G for industry applications. The focus is on brownfield investments, counting on the production planner to be able to provide data regarding the status quo of the application both from a technical and economic perspective. The framework implements a 4-step approach with the following steps:

1. Requirement check: In this step, the production planner selects a use case as well as the network requirements from the end-user point of view
2. Goal definition: Technical and economic goals are determined.
3. Data acquisition: In this step, the production planner enters relevant data of the use cases to be analyzed.
4. Process evaluation: Finally, the evaluation result of the technical and economic potential of 5G for the users' process is given.

Based on literature research and interviews with industry experts, this framework has identified seven industry goals. These are illustrated in Figure 3 and their definitions are further explained below [Lap14, ISO14a, ISO14b]. Flexibility describes the ability to process many parts within the manufacturing system with minimum engineering effort and change over time. Mobility describes the ability to move objects on the factory shopfloor. Productivity measures the output per unit of input over a specific period of time and therefore denotes the production efficiency. Quality rates the degree to which the output of the production process meets the requirements. Safety is the ability of a system to protect itself and the operator from harm or accidents. Sustainability describes the level to which the creation of manufactured products is fulfilled by non-polluting processes that conserve energy and natural resources. Utilization is the ratio of actual machining time compared to the theoretically available time. All use cases implemented in the trials have been analyzed with respect to these industry goals. The details of the framework are described in [5GS21-D120].

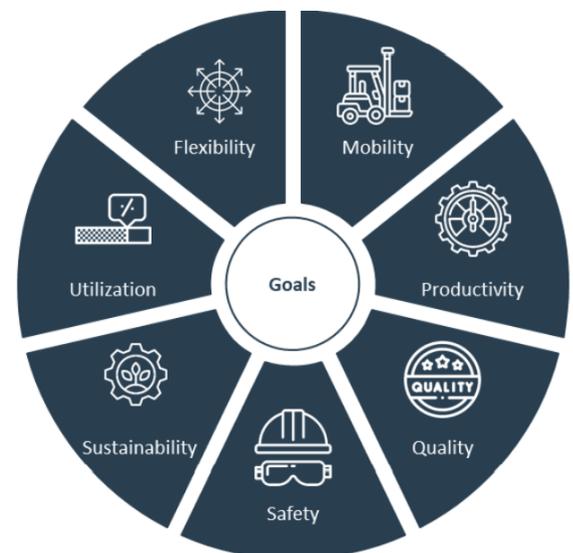

*Figure 3: Industry goals*

## ROLES, STAKEHOLDERS, AND BUSINESS RELATIONSHIPS

The vertical industries see 5G non-public networks (NPN) as a competitive advantage regarding wireless-based communication. Mobile Network Operators (MNO) are getting involved in provisioning 5G services for industrial customers, whose requirements may differ from the general public. More than any other cellular technology, 5G has initiated a shift towards a new ecosystem. New stakeholders have emerged or have diversified their activities to offer NPN design, integration, and operation services, and share liabilities, costs, value and profit. 5G-SMART has investigated how the relationships can be built between Industrial parties, MNOs, and other third parties and which value MNOs can bring to this ecosystem. The analysis is performed by defining relevant relationship models and then analyzing these based on a list of criteria identifying the main challenges to be addressed by the different stakeholders to fulfill the industrial end-user's needs and facilitate business relationships. The following categories for key-value creation criteria have been considered to analyze and compare relationship models from a business perspective:

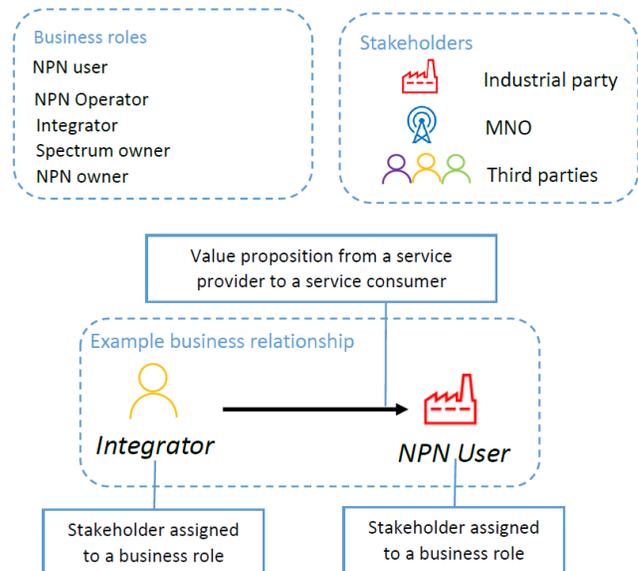

Figure 4: Roles, stakeholders and relationships

- Techno-related criteria to comply with performance requirements for industrial services,
- Business-related criteria facilitating the development of industrial activities,
- Technics- and business-related criteria towards NPN coverage extension and scalability,
- Security and confidentiality criteria to control risks and ensure data sovereignty,
- Economic criteria to assess needed CAPEX and OPEX investments.

Figure 4 illustrates the roles, stakeholders and relationships explored in 5G-SMART. They are further explained in [5GS21-D130]. Understanding and specifying the needs of the NPN User is key to targeting the right NPN deployment option and developing profitable business with involved parties.

The work in [5GS21-D130] contributes with an analysis on organizing who does what, understanding the strengths and weaknesses of each model, and evaluating the balance between the required investment and the potential enhancement of the overall efficiency (e.g., increased flexibility, optimized maintenance, etc.), security, etc.

| KEY RESULTS AND FINDINGS | <ul><li>There is no deployment and operation model, for which all criteria can be optimized, but a trade-off has to be found.</li><li>For industrial players it is important to develop a clear strategic direction of what to do with the 5G NPN and how much control or customization is required.</li><li>MNOs need to construct a suitable portfolio of offerings including both complex fully customized NPN deployments as well as standardized cost-efficient solutions.</li><li>There is an emerging possibility for third parties to support or complement MNOs, both in the design phase and in the operating phase.</li></ul> |
|---|---|

RELIABLE WIRELESS 5G CONNECTIVITY

Wireless connectivity is increasingly becoming a necessity for business-critical services in industrial processes, such as those related to assembly lines and other modes of production. However, the specific needs and requirements can differ greatly between the different industries. The communication network to be used has to be tailored for each particular deployment. While doing so, a large variety of aspects must be considered, e.g., characteristics of the use case requiring wireless connectivity, the overall business case, spectrum regulations, characteristics (relevant for deployment) of the industrial site, and so on.

A foundation for a wireless network is the availability of radio spectrum. 5G supports a wide range of radio spectrum bands, typically licensed to mobile network operators over a whole country or a region. In addition, it is possible in several countries to obtain spectrum usage rights for local 5G networks, for example, directly by a factory owner. Some examples of 5G spectrum bands often considered for local industrial 5G networks are around 3.5 GHz and 26 GHz. Several spectrum options can be used within a local 5G NPN, and an overview of spectrum options is provided in [5GS20-D140] and [Nor22].

A first step in planning a non-public industrial 5G network is to identify the different radio network deployment options so that it can provide the required communication services for smart manufacturing. 5G-SMART has identified and evaluated different radio network deployment options for smart manufacturing, to provide an overview of any desired industrial 5G scenario or service. It has been identified which input information is needed to select the most feasible deployment options, and the feasibility of several deployments has been analyzed. Moreover, the impact of spectrum options available for the stakeholder deploying and operating a non-public network has been discussed. A wide range of system-level simulation studies has been performed assessing the technical performance of the identified deployment scenarios. The detailed results from the discussions, analyses, and performance evaluations related to the different radio network deployment options are summarized in [5GS20-D140] and [5GS22-D150]. The capacity and required spectrum for smart manufacturing radio network deployment options have been investigated via radio network simulations. A complete trade-off analysis has been made for different bands, bandwidths, user density, and resource allocation. This analysis has shown that the system limits of capacity can come from the limitation of resources and, in some cases due to interference or propagation conditions.

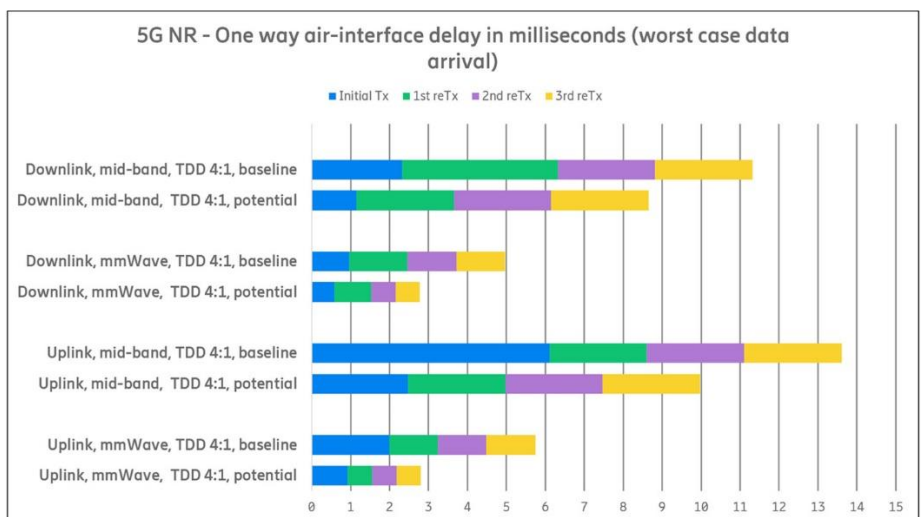
*Figure 5: 5G NR one way air-interface delay*

For example, Figure 5 gives an impression of how fast 5G New Radio (NR) can deliver data over the air, with and without retransmissions, and for different configurations. The results are based on typical processing delay values. Widely used in 5G networks, a time-division duplex (TDD) split of 4DL:1UL (DDDSU) slots is assumed for the resources into downlink (DL) and uplink (UL) in these spectrum ranges. More details on the assumptions for the 5G NR system for Figure 5 can be found in [5GS22-D150].

5G radio transmission is based on electromagnetic waves that propagate from the 5G transmitter to the 5G receiver and other locations where other equipment may be located. At such places, the 5G signal may create some interference. Coexistence describes the impact of mutual interference among multiple systems on each other. 5G-SMART has empirically assessed the coexistence between a wide-area outdoor network and an indoor NPN operating in the same channel. The results can be summarized as follows:

- Under some conditions, the transmission in the outdoor network can create interference to the indoor NPN and, therefore, increase the transmission latency of the indoor NPN. Such interference can appear in a (practically uncommon) situation, when an outdoor User Equipment (UE) is located right outside an unshielded factory window. This can lead to an increased NPN latency, which may be very small for the median values but can impact the 99th and 99.9th percentile of the latency distribution, in particular if the two networks use un-synchronized TDD configurations with substantial cross-link interference over a very short distance.
- Such cross-link interference can be avoided by an indoor TDD configuration that avoids Downlink (DL) transmission slots during Uplink (UL) transmission slots of the outdoor network while still allowing for a separate TDD configuration for the NPN.
- When the outdoor UE is located further away from the factory wall, the indoor NPN interference becomes negligible, independent of the used TDD pattern.

In general, from the different simulation studies performed within 5G-SMART, the following key observations could be made:

| KEY RESULTS AND FINDINGS | <ul><li>Local industrial 5G networks can be built based on multiple spectrum options. Mobile network operators have spectrum licenses on which local non-public networks can be deployed. In several countries, obtaining local spectrum licenses for industrial 5G networks is possible. Different spectrum options can be combined in a 5G network deployment.</li><li>Local 5G networks can be deployed in different options, for example, the number and types of antennas installed or the configuration of the time-division duplex pattern. Various 5G deployment options achieve different performance and capacity, and a suitable 5G deployment must be selected that matches the intended use cases.</li><li>In typical situations, a local indoor industrial 5G non-public network can be expected to coexist with an outdoor (public) mobile network using the same channel without significant mutual interference well.</li><li>5G includes capabilities for ultra-reliable low-latency communication (URLLC), which can support time-critical services with a need for guaranteed message delivery within a specific latency bound.</li></ul> |
|---|---|

# 5G deployments and performance at the trial sites

Three testbed sites at shopfloors with different characteristics have been built up in the project:
- Kista trial site, Sweden
    - The Kista trial site is located inside the Ericsson smart factory in Kista, outside Stockholm, and consists of a testing area of around 50 sqm.
- Aachen trial site, Germany
    - The Aachen trial site is part of the 5G Industry Campus Europe and is located on the shopfloor of the Fraunhofer IPT. The 2700 sqm large shopfloor is equipped with various kinds of machine tools from different vendors, thus reflecting a production landscape representative for many manufacturing companies. Both indoor and outdoor networks are deployed at the Aachen trial site.
- Reutlingen trial site, Germany
    - The Reutlingen trial site is the cleanroom factory floor of around 8000 sqm inside a Bosch semiconductor factory in Reutlingen. The shopfloor is characterized by narrow corridors, a high density of machines, and cleanroom conformity, which implies a very low number of particulates, a prerequisite for semiconductor manufacturing.

For all of the 5G deployments at the three trial sites, the choice has been made to design real need-based and future proof deployments fulfilling the use cases' requirements and in a wider realistic setting. A key learning from analyzing the requirements and KPIs of the use cases has been the identified need for a reliable, low-latency, high-performance wireless communication infrastructure as provided by 5G in future factories. Moreover, several of the use cases in Table 2 benefit from seamlessly moving devices, machines, and assets between factory halls, factory sites, and even countries. This possible need for mobility and outdoor wide-area connectivity is supported by 5G. The use cases implemented and demonstrated at the trial sites are only single proof points. Scaling from one use case to several hundred use cases and devices on the factory floor is needed, and 5G is expected to support this scaling. Introducing wireless connectivity on the factory floor increases the potential points of vulnerability, regardless of the network. 5G security builds on and significantly augments an underlying approach to security that is well established in the standards organization developing protocols for mobile telecommunication (3GPP).

### ARCHITECTURES, SPECTRUM, AND PERFORMANCE

Table 2 shows the deployments that were undertaken at the three different trial sites.
In the testbeds the two different architectures of 5G were considered: 5G Non-standalone Architecture (NSA) and the Standalone Architecture (SA). In NSA both 4G LTE and 5G NR were used, whereas in SA

| Trial site | Trial System | 5G Architecture | 5G Frequency Band | 4G Frequency Band | 5G Bandwidth [MHz] |
|---|---|---|---|---|---|
| Aachen | Midband | NSA | 3.6 GHz | 2300 MHz | 100 |
| Aachen | Midband | SA | 3.6 GHz | - | 100 |
| Kista | Highband | NSA | 28 GHz | 1800 MHz | 200 |
| Reutlingen | Midband | SA | 3.6 GHz | - | 100 |

Table 2: 5G deployments at the trial sites

5G NR acts as dedicated radio access technology for both control and data traffic, and no anchor needs to be established to a 4G network. In order to operate the 5G system at the trial site, an application for a local spectrum license had to be made. For the testbeds located in Germany, the possibility of applying for local spectrum licenses in the 3.7-3.8 GHz range were exploited. The deployments established at the 5G-SMART trial sites are all Non-public Networks (NPN). They are all deployed fully on the organization's defined premises. The architectures and frequency bands for the different sites were selected to enable the evaluation of use cases at different deployment options. The Ericsson Radio Dot System (RDS) was deployed at all testbeds, specifically designed for indoor deployments.

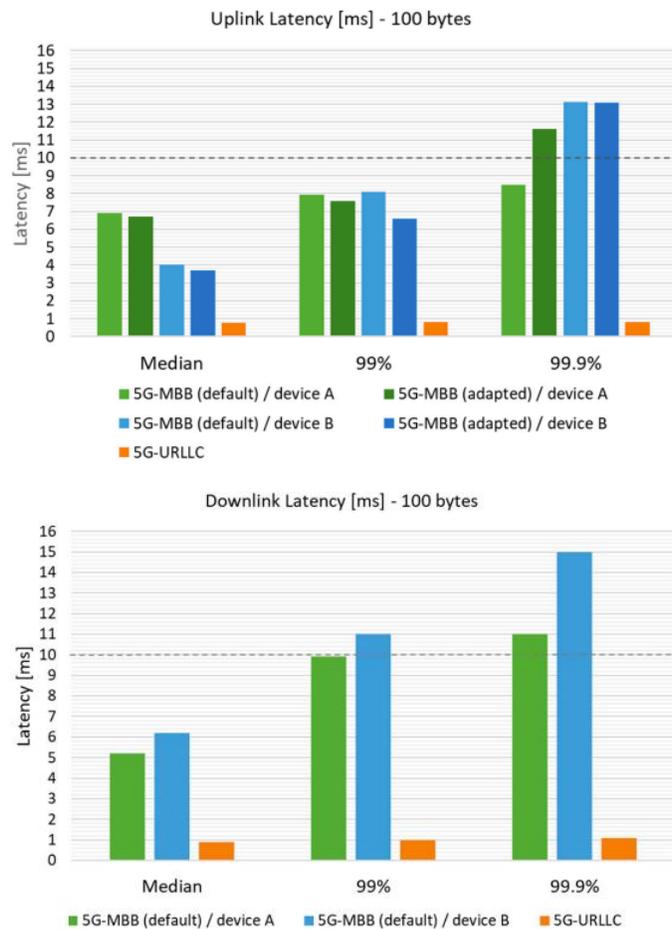

To characterize the expected performance, various systematic latency and reliability measurements were performed at the trial sites, over the air and in live networks. The results are published in [Ans22]. Measurements have been undertaken for various message sizes, different devices, and network configurations. In Figure 6 as an example from the measurements, the uplink and downlink latencies are shown for a message size of 100 bytes for different 5G devices and network configurations. The small message size is typical for industrial control applications. As can be seen, the 5G system deployed today can achieve low average latencies of 3.7-6.9 ms in uplink and 5.2-6.2 ms in downlink. The $99^{th}$ and $99.9^{th}$ percentile values indicate which latency can be provided for 99% and 99.9% of the transmissions. As can be seen, today's 5G systems are primarily designed and configured to provide good average latency performance while providing stringent guarantees on the latency bounds has not been the focus.

*Figure 6: UL and DL latency measurements, midband*

| KEY RESULTS AND FINDINGS | <ul><li>Today's 5G latency performance significantly depends on packet size, transmission direction (uplink or downlink), network configuration, and the 5G device's design and capabilities.</li><li>The requirements for very low latencies can be achieved with 5G-URLLC, as required in some of the most stringent industrial IoT applications.</li></ul> |
|---|---|

# Kista testbed: 5G-enhanced industrial robots

In a continuous effort to boost productivity, industrial automation and smart manufacturing are seeking to embrace state-of-the-art technologies from the domains of, e.g., Internet of Things, artificial intelligence, and robotics. Industrial robots have a key role in smart manufacturing, relieving human workers of highly repetitive tasks and improving their safety. The manufacturing landscape is also witnessing an increasing number of production lines in which stationary robots collaborate with human workers or are supported by mobile robots for transporting materials and goods.

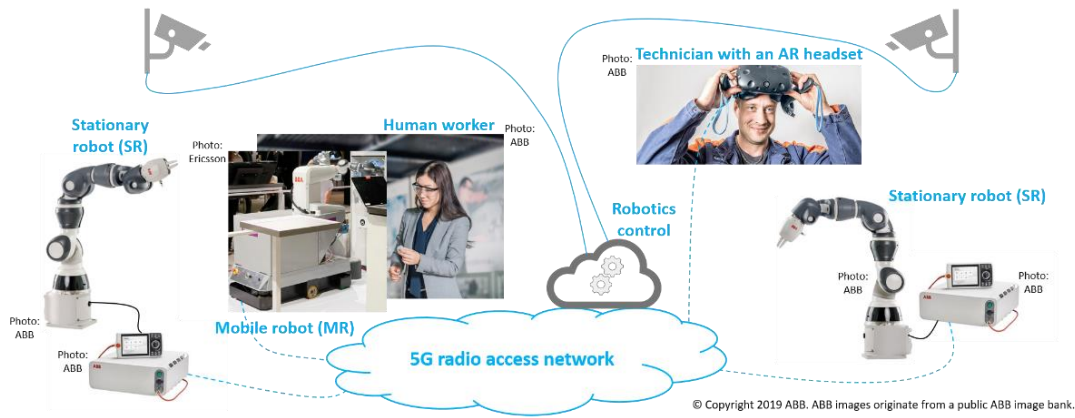

*Figure 7: Use case illustrations, Kista trial site*

At the Kista trial site, the testbed demonstrates, evaluates, and validates 5G capabilities for 5G-enhanced industrial robotics inside the Ericsson smart factory. Robotics is a vital part of modern manufacturing. 5G wireless communication and edge-cloud computing are two technical trends that may disrupt how industrial robots are deployed and used in the future. The 5G-enhanced industrial robotics testbed validates the novel design of industrial robotics, where part of the robot control is moved from the robot hardware to a central location, e.g., a control room in the factory. This puts stringent requirements on 5G in terms of reliable and low-latency communication for connecting the robot to the controller. The use cases investigated are listed below and illustrated in Figure 7.
- Use case 1 on 5G-connected robots and their collaboration,
- Use case 2 on machine vision-supported real-time human-robot interaction, and
- Use case 3 on advanced visualization for the factory floor.

These use cases bring several advantages: The control hardware of industrial robots can be simplified and occupy less space on the shopfloor. This is achieved by moving control functionality into an edge cloud. Reducing the cabling with wireless connectivity improves the flexibility of redesigning the factory floor. 5G wireless connectivity allows increasing the number of mobile robots in a factory, which can take over more diverse tasks in a flexible production process.

The testbed setup incorporates two YuMi® stationary robots by ABB to realize autonomous pick and place of an object, a mobile robot platform with Light Detection and Ranging (LiDAR) sensors, used for implementing functional aspects of material transfer handling, a smartphone to demonstrate new ways of robot programming based on human-robot interaction, video cameras (Azure Kinect and Intel RealSense) to implement machine vision features of object detection and localization, and a Magic Leap Augmented Reality (AR) headset to demonstrate advanced features of visualization for the factory floor.

USE CASE: 5G-CONNECTED ROBOTS AND THEIR COLLABORATION

This use case focuses on a prevalent task in manufacturing, namely the scenario of autonomously transferring material and other items between production lines. The task has been realized by relying on the collaboration between different types of robots, namely a mobile robot and a stationary robot arm. In the setup implemented in the testbed, the mobile robot transfers an object between workstations with ABB's robot arms. This task involves (1) navigating the mobile robot in an environment shared with human workers and other possible obstacles, (2) machine-vision-assisted docking of the mobile robot to a precise pose next to each of the robot workstations, and (3) vision-assisted execution of the object pick-and-place operation by each of the stationary robots to grasp and move the object either from or to the mobile robot platform. All these operations are controlled from an edge cloud and over 5G.

Strict performance-related requirements on the communication must be fulfilled to realize the use case. Table 3 lists the requirements for the functionality of motion planning of the stationary and mobile robots. A more detailed analysis of the requirements can be found in [5GS20-D110], while details of the use case implementation are described in [5GS22-D220].

| Communication stream | Communication service availability | Maximum end-to-end latency | Average data rate |
|---|---|---|---|
| Motion planning for stationary robot | ≥ 99.99% | < [5-40] ms | DL < 1Mbit/s<br>UL < 1Mbit/s |
| Motion planning for mobile robot | ≥ 99.99% | < [10-50] ms | DL < 0.5Mbit/s<br>UL < 0.5Mbit/s |

*Table 3: Use case performance requirements*

The feasibility of the use case implementation over 5G has been validated in various validation scenarios and towards different KPIs. As an example, an important KPI for robot navigation, docking, and pick-and-place operation, is the completion time. The completion time is, for instance, the time it takes a navigation engine to plan and control the motion of a mobile robot along the desired path between start and goal positions. A low completion time directly contributes to increased productivity in the factory. Detailed documentation of the results can be found in [5GS22-D230]. The key results and findings are listed below.

<table>
<tr><td rowspan="2">KEY RESULTS AND FINDINGS</td><td>

- The tests and measurements for the collaborative tasks investigated in the testbed validate the feasibility of implementing mobile robot navigation, mobile robot docking, and stationary robot arm's pick-and-place planning from an edge cloud and having communication running over 5G.
- While functional design and implementation of the use cases were possible over a 5G network based on MBB services, i.e., without functionality for, e.g., bounded latency, several features have been identified as beneficial or required in the future. These include quality of service, bounded latency, time synchronization, etc.
- Selection of 5G devices for a commercial deployment will be critical, both regarding their general performance and supported integration features.
- Communication protocol solutions that are more adapted for use over wireless networks should be considered for future applications.

</td></tr>
</table>

## USE CASE: MACHINE VISION-SUPPORTED REAL-TIME HUMAN-ROBOT INTERACTION

Workers in a factory must have safe and efficient ways of interacting with machines and robots. An advanced approach to robot programming has been implemented in this use case. A human worker (e.g., a commissioning engineer) instantiates the use case by programming a stationary robot arm to execute the object pick-and-place operation autonomously after a lead-through demonstration. Instead of manually writing a robot program, the worker mimics grasping and moving the object through an application on a mobile device (smartphone), which the stationary robot then executes. In order to operate, the robot needs to learn two tasks: trajectory generation and object pick/drop. Such a scenario illustrates the potential of advanced means of human-robot interaction on the factory floor. This type of interaction also allows non-expert users to program robots, thereby increasing the adoption rate of robotics in new domains. Figure 8 illustrates the use case.

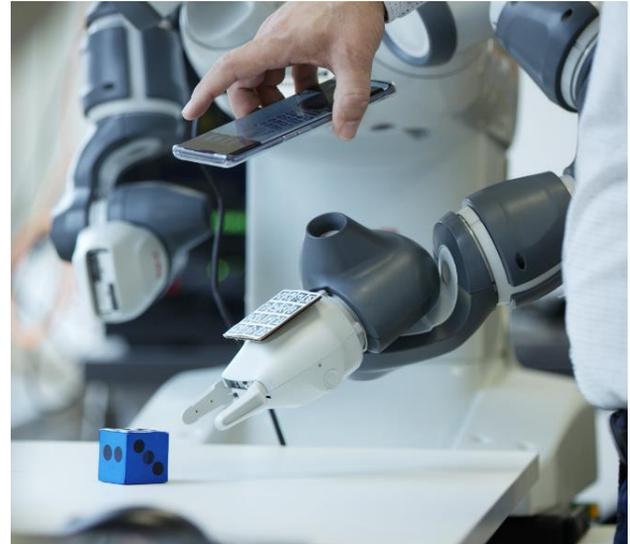

*Figure 8: Lead-through teaching*

The lead-through teaching via 5G use case was evaluated in validation scenarios both towards the latency requirements needed to execute the lead-through teaching operation and towards the user experience. Figure 9Figure 9 illustrates the two-way latency statistics for lead-through teaching as an example from the validation tests. The mean value of the round-trip-time (RTT) from the application in the smartphone to the edge and back is measured, together with the $90^{th}$ and $95^{th}$ percentiles of the distribution. The variation is due to different amounts of data sent from the smartphone and its transmission speed, where also the different angles of holding the smartphone camera during the teaching have an influence. More evaluation details can be found in [5GS22-D230].

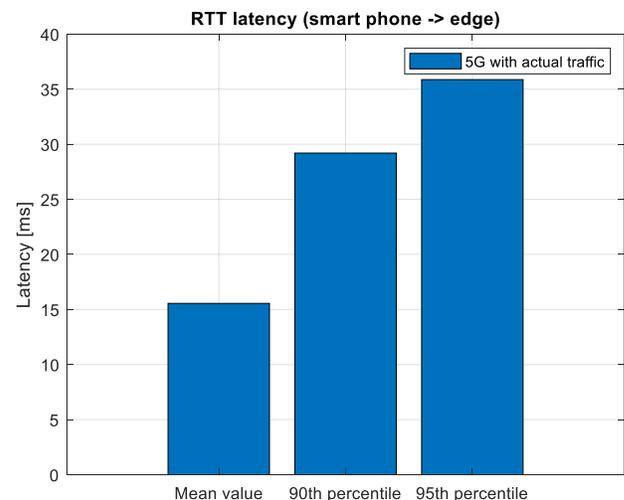

*Figure 9: Two-way latency statistics for lead-through teaching*

**KEY RESULTS AND FINDINGS**

- The feasibility of programming a robot arm by lead-through teaching via 5G has been demonstrated, with the 5G network supporting the application well.
- Novel robot programming techniques such as lead-through teaching improve the flexibility to re-configure the robots for different tasks and open the possibilities for non-expert users to interact efficiently and program robots.

## USE CASE: ADVANCED VISUALIZATION FOR THE FACTORY FLOOR

The AR visualization of robot status shows the potential of AR technology to supervise industrial robots and their every-day operation, to increase the work efficiency of human technicians on the factory floor, and to make human-machine interaction simpler. It also showcases the usability of 5G networks to enable low-latency exchange of data and commands between mobile AR-using technicians and industrial robots.

This use case is characterized by employing AR-based means to manipulate robot motion remotely and visualize operational robot information efficiently. Machine vision support is used to detect and distinguish between different industrial robots.

The implemented scenario is shown in Figure 10. It contains the two single-arm YuMi® robots, for which information visualization and control is performed via the AR headset. It should be noted that the test setup includes two 5G communication legs over the air: both the AR headset and the robot controllers are connected via 5G, each with their own User Equipment (UE).

The use case has been validated in two test scenarios, which involved reading of AR robot data from one robot arm or alternating information retrieval and visualization between the two robots over 5G. The evaluation details can be found in [5GS22-D230]. Below the key learnings and findings are summarized.

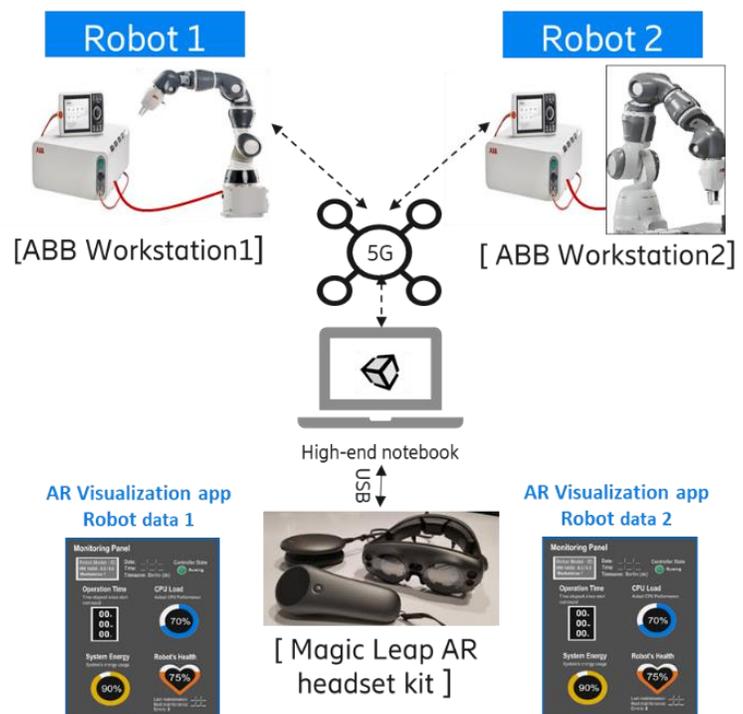

*Figure 10: AR use case*

**KEY RESULTS AND FINDINGS**

- The AR use case can be implemented over 5G and applied to maintenance and monitoring on the factory floor. The latency performance can be considered sufficient for the applicable scenarios.
- Subjective evaluation tests showed that knowledge of related AR technology and/or 5G does not affect the overall quality of user experience. Moreover, people over 50 years of age did not have issues regarding the quality of experience and technology. However, it was observed that younger people (below 30) seem to be more critical in terms of the quality and experience and if the technology is ready to be applied in the given scenarios. These findings may impact the way new technology is introduced on the factory floor.
- AR visualization of this type is expected to contribute positively to industry goals such as productivity, sustainability, and utilization.

# Aachen testbed: 5G for enhanced industrial manufacturing processes

The Fraunhofer IPT trial site in Aachen deals with the application of 5G to industrial manufacturing use cases and addresses various specific aspects of this production area. On the one hand, it focuses on multiple types of process monitoring. On the other hand, it supports the condition monitoring of assets in the factory, such as machines and its infrastructure.

Due to the fact that specific applications require wireless sensors (e.g., 5-axis milling) and the demand for tight timing conditions, new solutions for future production are needed. Existing wireless communication standards lack performance in terms of latency and data throughput. Bluetooth, on the one hand, delivers a rather low latency in the single millisecond range, yet it suffers from communication range and data rate. It, therefore, is limited to point-to-point connections, typically inside machines. Because of that, it generates increased efforts to integrate Bluetooth sensors into production IT. On the other hand, Wi-Fi communication shows an acute limitation in terms 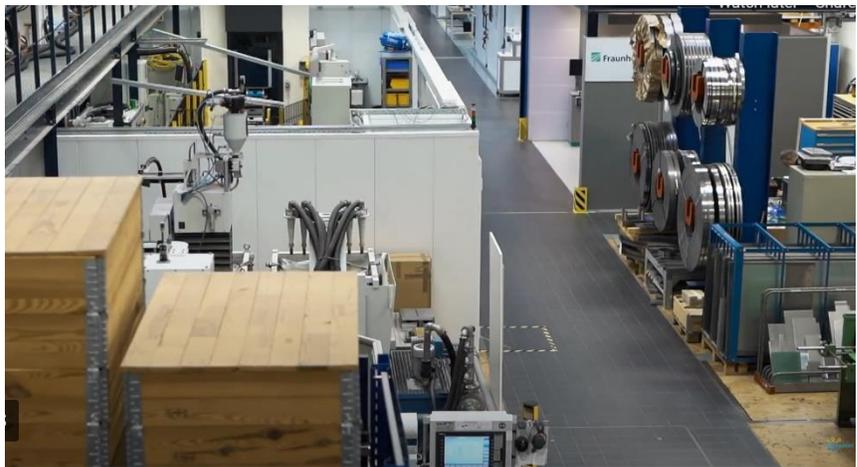 of jitter-free communication. This is critical for immediate reactions, e.g., stopping machining upon tool breakage detection or collisions of the machine spindle. Acoustic emission sensors typically have sampling rates in the MHz range and can generate data output rates of up to several Mbit/s depending on the application. In conclusion, low signal transmission latency and high data throughput are required. While all the above mentioned technologies cannot fulfill high requirements of timeliness, reliability, data rates, scalability, and availability, 5G is seen as a key enabler in monitoring applications to manufacturing.

Besides that, wireless solutions can be easily retrofitted compared to wired solutions, given that the wireless solution does not sacrifice the performance delivered by the wired solution. Here, 5G is a potential candidate for the connectivity of retrofit manufacturing equipment.

In the Aachen testbed, considerable attention has been put into validating the use cases in relevant scenarios in a realistic industrial manufacturing environment. Instead of an artificial lab, which does not represent a real production environment, use cases have been demonstrated and validated on an industrial shopfloor. Specific parameters like end-to-end latency have been benchmarked and compared against the current state-of-the-art, which in most cases is a wired off-the-shelf solution with analog sensors and electronic devices, typically located in the electrical cabinet of a machine and a real-time operating system for the measurement data processing. The performance values resulting from this benchmarking can then be used to calculate additional KPIs.

USE CASE: TOOL CONDITION MONITORING USING A 5G ACOUSTIC EMISSION SENSOR

In this use case, an acoustic emission (AE) sensor has been developed and integrated into a 5-axis milling machine, to monitor the condition of cutting tools. Acoustic workpiece monitoring is a technology that uses AE sensors to collect relevant data for the monitoring system. AE sensors are widely applied for monitoring cutting processes. In this use case, they are used for monitoring tool wear, detecting tool breakage, and detecting the collision of the machine spindle. Timely detection of any of the above disturbances is highly desirable as it allows an intervention into the process to optimize the fabrication process and reduce the production costs due to decreased failure rates.

The setup implemented is illustrated in Figure 11: a wireless AE sensor with a 1 MHz sampling rate has been integrated with a 5G device. The sensor is connected to the workpiece during the machining process to provide measurements from the machine to a monitoring unit, e.g., located in an edge cloud. The raw signals are preprocessed on the wireless device and then transmitted via 5G as User Datagram Protocol (UDP) packets with a data length of 1024 bytes and a frequency of 1 ms. The measurements are analyzed inside the Monitoring Unit – Genior Modular (GEM)[1], and the observations are fed to the machine control to steer the machining process. The entire loop of acoustic emission measurements, data collection and analysis in the monitoring unit, and adaptively steering the machining process from the machine control needs to take place in real-time.

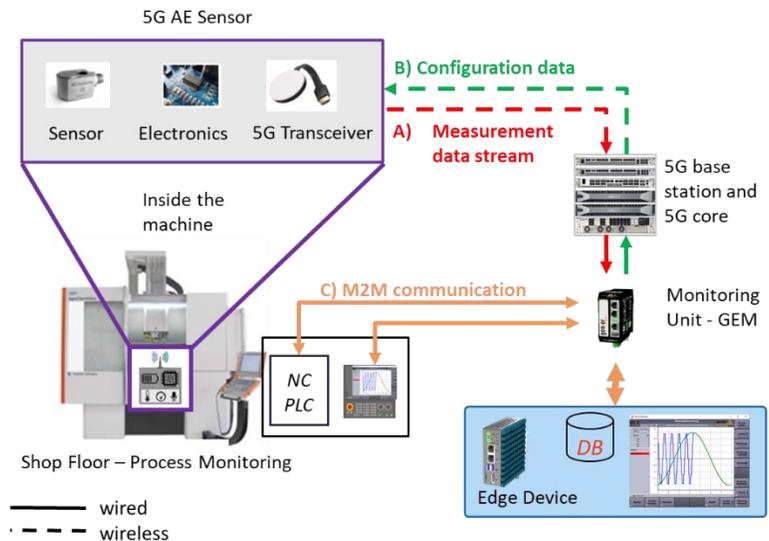

Figure 11: Acoustic Emission sensor test setup

The use case has been evaluated in different validation scenarios as well as towards the previously introduced industry goals, see [5GS22-D340]. For example, improved tool breakage detection addresses the industry goal of more sustainable production. In milling processes with small tool diameters where the cooling lubricant is used, tool breakage can occur during the process without the machine operator noticing. This causes the milling process to continue unnecessarily, wasting machine energy and cooling lubricant. 5G sensor-based detection of tool breakage allows the machine to be stopped immediately, saving energy and resources and improving the sustainability of the process.

| KEY RESULTS AND FINDINGS | <ul><li>The AE sensor system can be successfully realized over 5G. Several benefits are observed concerning the industry goals of flexibility, productivity, utilization, and sustainability.</li><li>A limitation of the current implementation can be seen with respect to energy consumption. A need for the development of 5G UEs with reduced power consumption to allow reasonable battery runtimes or sizes has been identified.</li></ul> |
|---|---|

---

[1] https://www.marposs.com/eng/product/tool-and-process-monitoring-system-2

USE CASE: PROCESS MONITORING USING A 5G MULTISENSOR PLATFORM

This use case aims to address and solve the limitations of current sensor systems. The vision of the Multi-Sensor Platform (MSP) is to collect heterogeneous data from a variety of different sensors on the shopfloor, transfer these via 5G and aggregate them in a local cloud close to the shopfloor. The general concept can be seen in Figure 12: on the shop floor, multiple machines and workpieces as well as the infrastructure are equipped with MSPs, and are connected via 5G to the local cloud, where measurement data can be processed and stored. Some extracted information can then be fed back as process parameter adjustment or control to the machines. Sensors are tuned and orchestrated in the form of configuration data.

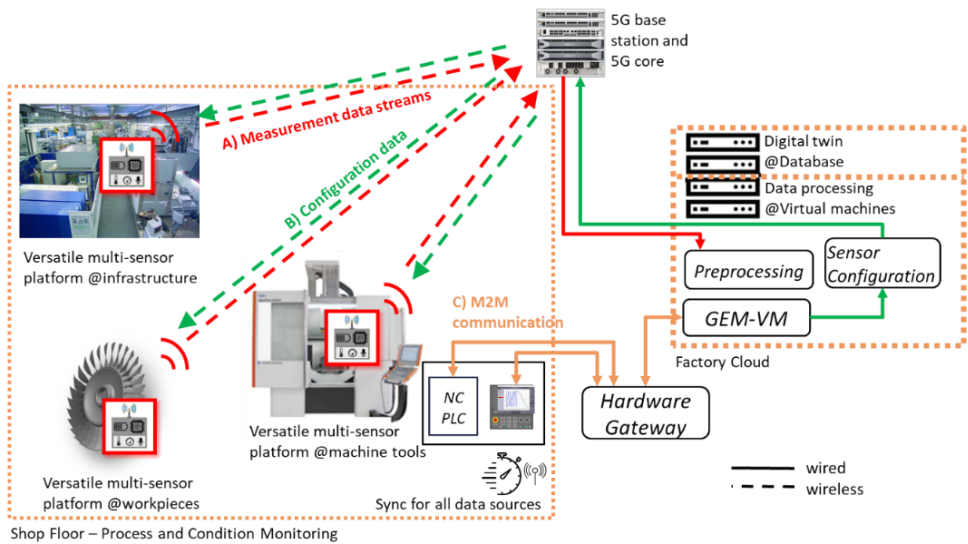

*Figure 12: Use case setup Multi Sensor Platform*

Many diverse physical quantities can be measured or sensed across a factory, relating to machines, workpieces, and the infrastructure as well. Each of those may have different requirements, especially regarding reliability and latency, which can be challenging. Critical process parameters in machining are, for example, accelerations or forces, which are an indicator of the unpredictable behavior of the workpiece to be machined. Chatter marks or tool deflection may result, leading to insufficient quality of the final product. To instantly react to such incidents, latency less than 10 ms may be required to adopt the machining parameters. The MSP has been evaluated in various validation scenarios. Detailed results and discussions of the Aachen trial site can be found in [5GS22-D340].

<table>
<tr><td>KEY RESULTS AND FINDINGS</td><td><ul><li>The MSP can be realized via 5G and will provide a valuable tool to rapidly ramp up new processes and achieve a high product quality in a short time, thus saving time and energy.</li><li>The 5G communication of the MSP can be successfully interfaced with state-of-the-art monitoring equipment.</li><li>Time synchronization over the 5G mobile network enables industrial users to extract ultra-high accuracy timestamps, as demonstrated with the CellTime™ solution integrated into the MSP.</li></ul></td></tr>
</table>

# Reutlingen testbed: 5G for semiconductor factory automation

CHANNEL MEASUREMENTS AND ELECTROMAGNETIC COMPATIBILITY TESTING

The semiconductor factory floor is a challenging environment in terms of radio propagation, not only due to its layout consisting of narrow corridors and high walls but also due to the large amount of reflective material/equipment. Therefore, attention needs to be given to providing reliable coverage throughout the factory. Factory floors are usually environments characterized as rich scattering with various tools and machines, which contribute to shadowing effects and are in different ways interacting with the radio signals. In addition to that, moving people and robots introduce dynamic changes in the environment. 5G-SMART has designed channel measurement campaigns to study the effects of these aspects, measuring signal properties, such as signal strength, impulse response, delay spread, etc.

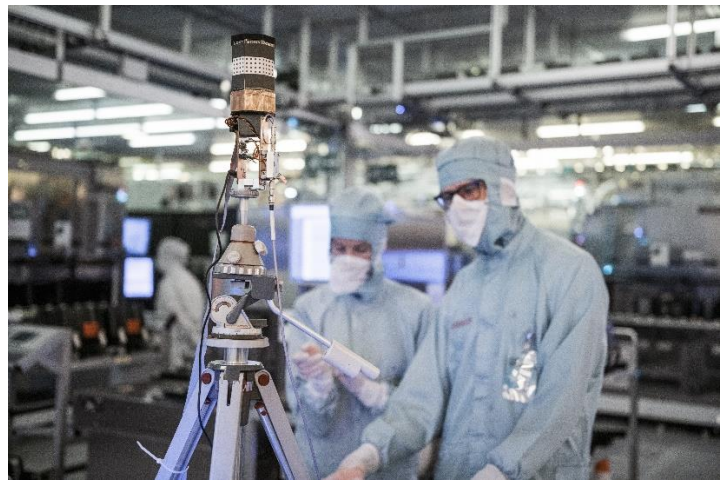

Different scenarios and locations have been determined to obtain a comprehensive picture of the radio propagation characteristics in the factory. The results show that providing coverage at midband frequency looks promising. Furthermore, this was confirmed during the use case execution phase, where no coverage issues were detected. 5G-SMART's deliverable D4.2 [5GS21-D420] describes the performed tests and analysis in detail.

A primary requirement to install 5G in industrial environments is that 5G is electromagnetically compliant with the industrial equipment on site. Therefore, it is crucial to make sure that 5G signals do not negatively impact the production processes. To address this issue, 5G-SMART has analyzed and evaluated the impact of 5G signals on semiconductor production using a test setup developed by the project. The setup uses a vector signal generator generating uplink or downlink 5G NR signals at 3.7 GHz with 20 MHz and 100 MHz bandwidths, which are amplified and transmitted via a horn antenna. The impact of the electromagnetic fields created due to the 5G signals has been investigated for different devices under test. Based on the Electromagnetic Compatibility (EMC) test results, it has been concluded that with the actually tested samples, deploying a 5G network in the final test and the sensor backend areas can be considered without any concerns, while the wafer test area cannot be considered unless actions are taken to ensure EMC. Details of the analysis and results can be found in [5GS21-D420].

| KEY RESULTS AND FINDINGS | <ul><li>Factory floors can be a very challenging environment in terms of radio propagation. Channel measurements show, however that providing coverage at midband frequency is also possible in very dense shopfloors.</li><li>In semiconductor factories, electromagnetic compatibility is a major requirement due to the sensitivity of the wafers.</li></ul> |
|---|---|

## USE CASE: CLOUD-BASED MOBILE ROBOTICS

This use case focuses on the feasibility, flexibility, and performance of wirelessly controlled mobile robots in a manufacturing shopfloor with 5G technology. Two autonomous mobile robots (AMRs) have been enhanced with 5G to demonstrate different aspects of AMR cloudification: A research AMR and a 5G-enhanced commercial AMR. In Figure 13, the functional architectures of a state-of-the-art commercial AMR (left side), the 5G-enhanced commercial AMR (middle), and the custom-built Research AMR (right side) are illustrated for comparison. As can be seen, more and more components are offloaded from the hardware device to the local factory cloud when moving from the commercial AMR towards the Research AMR. The latter one has been entirely built up and implemented by the 5G-SMART project with *all* software blocks in the cloud. Even the safety functions are executed in the cloud, but they are also kept onboard to be in line with the safety regulations of the factory. This means that, in principle, the entire intelligence of the research AMR is removed from the platform itself, reimplemented, and extended in a cloud-native manner to the edge cloud. The certified commercial AMR platform is connected over the 5G system to show the benefit of collaborative knowledge collected in the factory cloud (e.g., the two AMRs share in real-time a common map for trajectory planning and optimal shopfloor navigation). A detailed description of the selection of robot platforms, necessary sensors, and other equipment, as well as the implementation of the hardware and software architectures, is provided in 5G-SMART deliverable D4.3 [5GS21-D430].

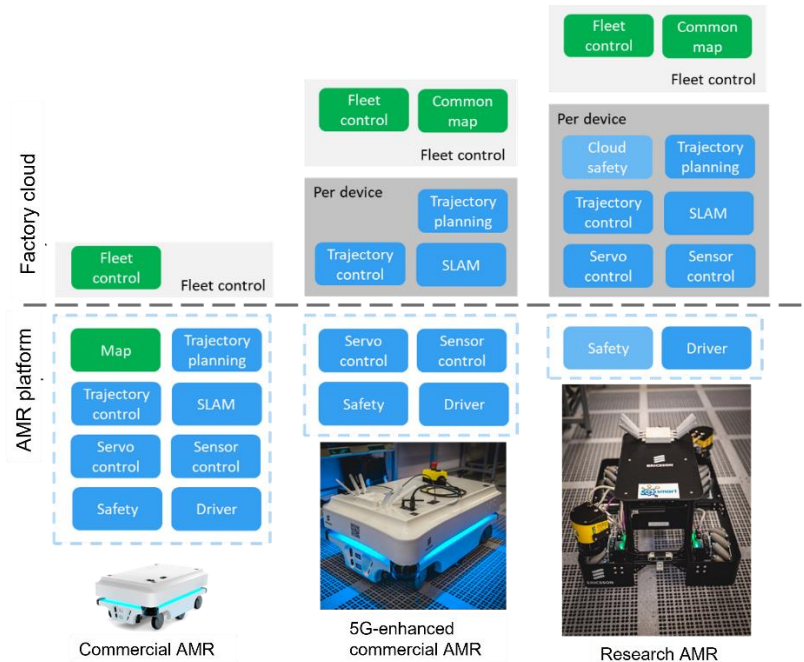

*Figure 13: AMR architectures*

The detailed documentation of the cloud-based mobile robots' use case validation and evaluation can be found in Deliverable D4.4 [5GS22-D440].

| KEY RESULTS AND FINDINGS | <ul><li>The deployed 5G network could support all of the tested validation scenarios without any deterioration in the performance of the robots. This means that cloud-controlled mobile robots can be successfully realized via 5G.</li><li>By using 5G and cloud technologies, novel collaborative control solutions can be enabled for autonomous mobile robots, e.g., optimizing the logistics in factories.</li><li>In the evaluations of the industry goals, a positive impact on efficiency, flexibility, productivity, mobility, and safety was found.</li></ul> |
|---|---|

USE CASE: TSN/INDUSTRIAL-LAN OVER 5G

This use case focuses on investigating and validating the applicability of 5G for transporting the traffic of Time Sensitive Networking (TSN)/industrial LAN (I-LAN) applications. Due to industrial applications' stringent requirements, all operational I-LANs are realized based on fixed (wired) communication networks. Limited flexibility for setting up new production lines or restructuring an existing production line, as well as complex and costly maintenance, are major drawbacks of the wired I-LAN realizations. In particular, this can be an issue given the recent trends for making the industrial environments as flexible as possible, e.g., smart factories of the future in the context of Industry 4.0. Introducing 5G can potentially reduce the cables and connectors' wear and tear for the mobile machines/controllers, resulting in reduced maintenance costs. Additionally, replacing the cables for communications between controllers and machines with 5G communications results in greater flexibility for implementing and adapting the industrial manufacturing infrastructure. Consequently, this can improve manufacturing productivity by reducing the time for setting up or customizing a production cell/line and improving the maintenance. Partially replacing fixed interconnections between TSN/I-LAN nodes with 5G mobile communications puts very stringent requirements in terms of latency and reliability on the communication system. Figure 14 shows the testbed setup implemented at the Reutlingen trial site. Two communication streams are investigated: the communication between Programmable Logic Controllers (PLC) and the communication of an industrial machine with its backend server. In order to validate the applicability of TSN/Industrial LAN traffic over 5G, diverse validation scenarios have been designed and tested for the applications in this use case. More specifically, the performance of these applications has been evaluated under various load conditions at the RAN by also investigating options of prioritizing different types of traffic. The detailed analysis of the TSN/I-LAN use case can be found in Deliverable D4.4 [5GS22-D440].

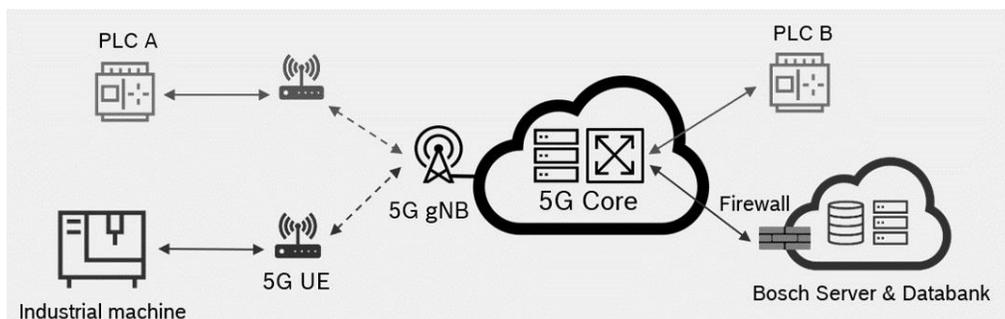
*Figure 14: Testbed setup Reutlingen trial site*

KEY RESULTS AND FINDINGS
- Controller-to-controller (C2C) applications can be realized via 5G, although the C2C communication performance over the wired Ethernet network outperforms that over the 5G network due to the shorter round-trip time.
- It has been demonstrated that without traffic prioritization for the C2C and machine-to-server communication, the background traffic may impact the controller traffic and the machine traffic in terms of increased delay. The conclusion is that, depending on the application, traffic prioritization may be needed.
- In the evaluations of the industry goals, a positive impact on flexibility, mobility, and sustainability was found.

# Technical features to optimize 5G for smart manufacturing

The 5G system must integrate with existing industrial communication technology. Integration and interworking of the 5G system with an existing industrial communication infrastructure is essential to enable communication services in a converged network. In order to ensure such interworking, 5G-SMART worked on enhancements of the 5G system including novel 5G technical features, network architectures and network management.

Three technical features were investigated to support the needs of advanced manufacturing use cases: end-to-end time synchronization, integration of 5G with Ethernet and time-sensitive networking (TSN), and 5G-based positioning. Figure 15 illustrates the use cases. The complete investigation can be found in [5GS20-D510] and [5GS21-D530].

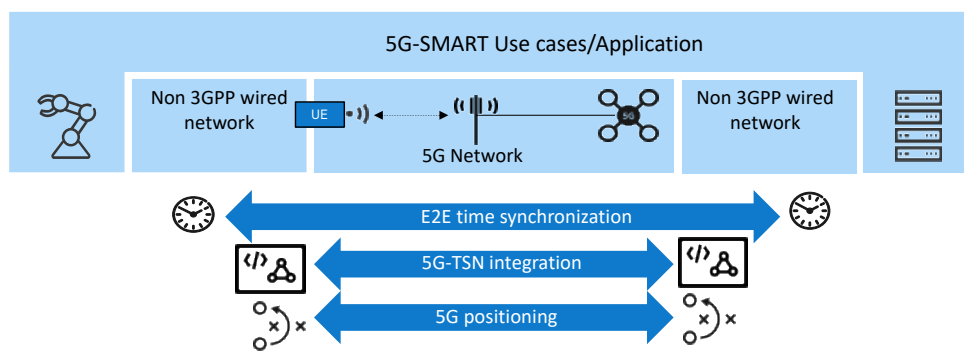

*Figure 15: Selected 5G features to support the 5G-SMART use cases, Source: Ericsson*

### 5G INTEGRATION WITH TSN

Bridged Ethernet is the general wired LAN networking technology used in factories, and larger subnetworks of the factory are interconnected via IP routing. Ethernet LAN is typically complemented with a wider set of Fieldbus technologies or real-time industrial Ethernet variants (such as PROFINET or EtherCAT), which can provide deterministic bounded latency performance within local network segments. This heterogeneous communication technology landscape is slowly changing and converging towards more open communication technologies. In this context, Time-Sensitive Networking (TSN) is foreseen as the open standards for wired deterministic low latency industrial communication in future. TSN provides deterministic services over IEEE standard 802.3 Ethernet wired networks. TSN standards can be applied to many verticals, including industrial automation. Moreover, DetNet, a deterministic transport solution to ensure bounded latency and low data loss within IP networks is a further potential technology for future smart manufacturing applications; DetNet is currently being specified by the Internet Engineering Task Force (IETF).

For the 5G system to be widely adopted in smart manufacturing it is a necessity to be able interwork with the industrial LANs, particularly Ethernet and Time-Sensitive Networking (TSN).

The following key technical enablers within the 5G system allow to interwork with industrial networks,
1. Ethernet connectivity support,
2. Time-sensitive communication support for TSN and DetNet integration,
3. Virtual network groups help to build a virtual Local Area Network (LAN).

With the support of the above technical enablers, today's 5G system (5GS) supports Ethernet and is compatible with IEEE standards relevant for industrial automation (IEEE TSN). This allows seamless interworking with existing Ethernet deployments, e.g., Local Area Network (LAN) integration and TSN deployment in industrial deployments.

Starting from Release 15, 5GS supports transport of the IEEE 802.3 Ethernet user plane traffic. 5GS also supports flooding and MAC learning as default forwarding mechanisms. The 5GS is modeled as a virtual Ethernet Bridge and is prepared for integration with the existing industrial network management system. From an external view, the 5GS can be seen as a virtual Ethernet bridge, and any network management system can interact and configure the 5GS similarly to an Ethernet switch.

In an industrial deployment, 5GS can have multiple 5G virtual Ethernet bridges, one per user plane function (UPF). The 5G core network binds the UPF instance ID (defined in TS 23.502 [TS20- 23502]) with the bridge ID.

Figure 16 shows the 5G integrated Ethernet network architecture, where 5G as a logical layer 2 Ethernet bridge exposes its capabilities to the existing network management system. These capabilities include bridge and port management information such as Link Layer Discovery Protocol (LLDP), Virtual LAN (VLAN), and other related IEEE 802.1Q configurations.

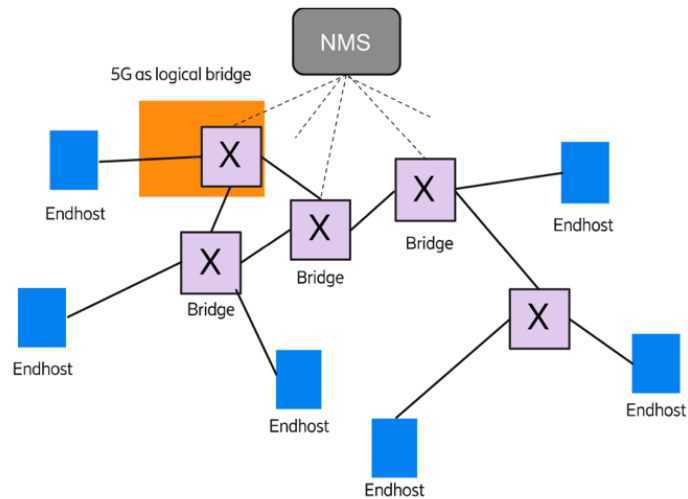

*Figure 16: 5G-integrated Ethernet network architecture*

From release 16, 5GS supports interworking with TSN-based networks. 3GPP has specified 5GS interworking through its support of the time-sensitive communication services. Figure 17 shows the integration of 5GS with the TSN-based Ethernet network. The 5GS provides Ethernet ports on the mobile device sides (Device Side TSN Translator DS-TT) and the network side (Network Side TSN Translator NW-TT).

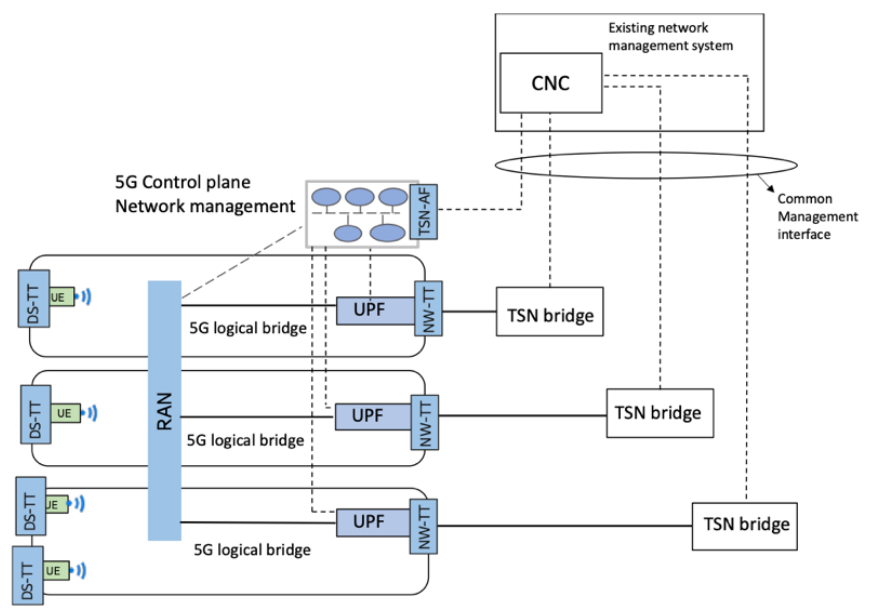

*Figure 17: 5G-TSN deployment model*

Ethernet and TSN stream communication is possible between any ports over the 5GS, including DS-TT to DS-TT communication (i.e., UE to UE communication via the UPF). The 5GS also specifies the TSN application function (TSN-AF), a control plane application function that interacts with the TSN Centralized Network Configuration (CNC). 5G bridge properties and 5G TSN functions are configured via the TSN AF. Ethernet/TSN frames are mapped to 5G specific QoS flows based on the Priority Code Point (PCP) according to IEEE 802.1Q standard. This way, the TSN-AF triggers the configuration of required capabilities within the 5GS based on the information exchanged with the CNC.

| KEY RESULTS AND FINDINGS | ▪ 5G can integrate with industrial networks and supports both IP and Ethernet communication. 5G supports time-sensitive communication allowing tight 5G integration with Ethernet Time-sensitive Networking (TSN), and in future also IP Deterministic Networking (DetNet). |
|---|---|

### END-TO-END TIME SYNCHRONIZATION

Time synchronization over the network is one important capability for industrial automation use cases, and is also one main building block for IEEE 802.1 TSN-based time-aware industrial networks. 3GPP has proposed new mechanisms in Release 16 in order to represent the 5GS as IEEE 802.1AS time-aware system in an integrated 5G-TSN network. In order to also cater for industrial use cases with highest demands on permitted time synchronization error; the 5GS has been specified to achieve time errors that can be bounded to less than 900 ns between the ingress and egress points of the 5GS.

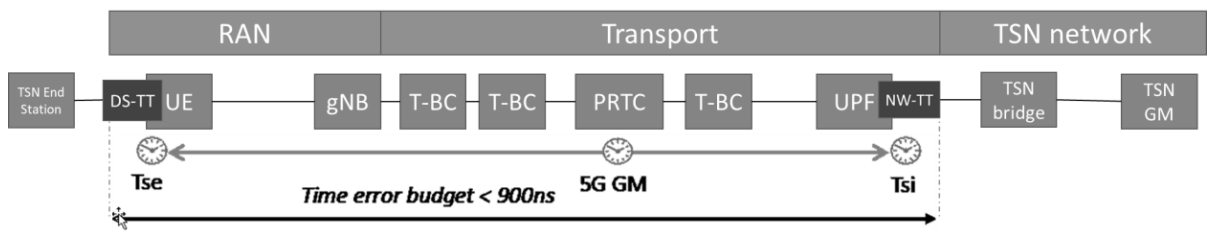

*Figure 18: Time error analysis*

The radio link between the base station and the user equipment (UE) is thereby identified as the major contributor to the time error of the 5GS, e.g. from the uncertainty of the estimated downlink propagation delay, which is used by the UE to adjust the 5G reference time. 5G-SMART has investigated and evaluated time synchronization of the 5GS for time synchronization. Key learnings and findings are listed below.

| KEY RESULTS AND FINDINGS | ▪ It was demonstrated by analytical analysis and simulations that among the two mechanisms proposed in 5G standardization for time error compensation of the radio link, the round trip-time (RTT) based propagation delay compensation method can achieve lower time error compared to Timing Advance based method. |
|---|---|

## POSITIONING

Reliable positioning with high accuracy is important for navigation and tracking of devices, AGVs, and persons in a large set of use cases in smart manufacturing. The performance of the 5G-radio frequency-based positioning of a device in a realistic industrial environment has been investigated and evaluated within 5G-SMART, with focus on the multilateration positioning methods time-of-arrival and time-difference-of-arrival. For the evaluation of positioning performance an industrial 5G deployment similar to the 5G-SMART trial site at Fraunhofer IPT was selected, for which both a statistical radio propagation channel model according to 3GPP was used, as well as a geometric 3D ray tracing model. The achievable positioning accuracy was obtained, and the impact of the choice of channel model on the evaluated positioning accuracy was determined. As an example of the evaluation, Figure 19 shows the CDF of positioning errors in the different scenarios corresponding to two different carrier frequencies (FR1 = 3.5 GHz, FR2 = 28 GHz), UE height levels (0.5 m and 1.5 m) as well as the bandwidth (100 MHz and 400 MHz). From these curves, the increase in bandwidth shows a pronounced improvement in positioning performance from m-level to dm-level. However, this improvement corresponds to the Line-of-Sight (LoS) positions, whereas the positioning error on the non-LoS areas is not improved by the increased bandwidth. According to this figure, in those areas the positioning error can be improved by increasing the UE height. Considering the 90% confidence level as a reference point, it is interesting to see that the FR2 scenario with 400 MHz bandwidth and a UE height of 1.5 m has demonstrated positioning errors below 30 cm, under the assumption of perfect time synchronization between base stations. As this solution requires no additional infrastructure, this is a promising result. For use cases that require even better accuracy, 5G-based positioning can be combined with other sensors, such as lidars, cameras, inertial measurement units, etc. More details of the analysis can be found in [5GS21-D530].

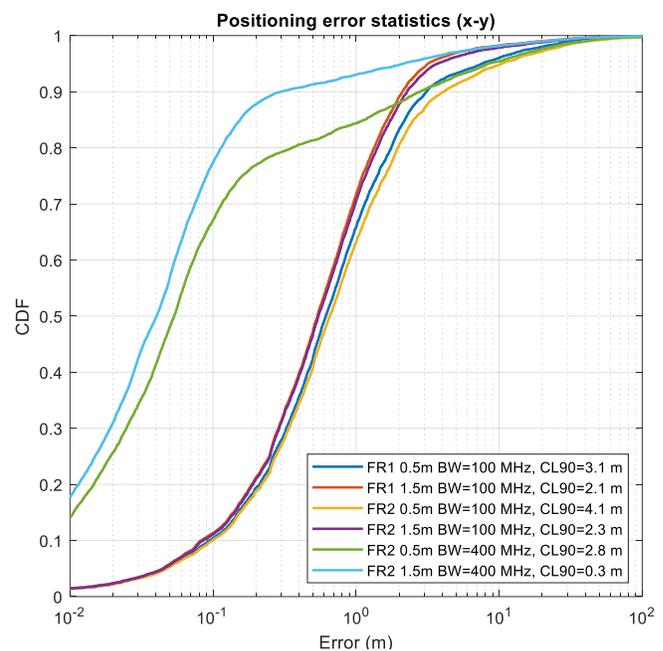

*Figure 19: Positioning error statistics based on geometric model-based evaluation. The 90% confidence level for each curve is stated in the legend.*

| KEY RESULTS AND FINDINGS | - The potential of 5G for Industrial IoT positioning look promising. The positioning performance depends on various factors and can be improved by ensuring good line-of-sight conditions and a large bandwidth.<br>- For many use cases the 5G system is sufficient to provide both communication and positioning. |
|---|---|

# Novel 5G network architecture options for smart manufacturing

5G for Industry 4.0 use cases needs to be integrated into an factory environment and provide low latency communication combined with high reliability and availability. This leads to considering various architectural options for mobile communication technologies that are different from conventional mobile broadband services. It is primordial to keep in mind that meeting the objectives of smart manufacturing applications involves an overall system perspective and, hence, includes not only the 5G network, but also the end system of the communication, i.e., the devices and the Edge cloud where applications are hosted. This section provides insights into network architectural aspects, including NPN, Edge computing, and device architecture.

### NON-PUBLIC NETWORKS

5G standardization has introduced new architectural options such as standalone non-public network (SNPN) and public network-integrated non-public networks (PNI-NPN) to enable smart manufacturing use cases. Considering the above smart manufacturing use case and its requirements, there is a need to investigate such NPN options, from both a deployment (shown in Table 4) and operation point of view. Furthermore, the key technical enablers related to the 5G integration with Ethernet, TSN, support for LAN type services, and network slicing need to be analyzed with respect to the deployment options of the 5G system in the smart manufacturing.

| NPN deployment options | Characteristic/details | 3GPP NPN type |
|---|---|---|
| Standalone NPN | All NPN functions are on-premises. NPN is a fully separate physical network from the Public Land Mobile Network (PLMN) with a dedicated NPN ID. However, dual subscription with the NPN and PLMN is possible. Access to PLMN services can be realized via an optional firewall connection and a roaming agreement. | SNPN |
| Shared RAN | The NPN is based on 3GPP technology with its own NPN ID. Only the RAN is shared with the PLMN, all other network functions remain segregated, also data flows remain local. It can be realized by:<br>• Multi-Operator Core Network (MOCN), where two or more core network entities are sharing base stations and spectrum<br>• Multi-Operator RAN (MORAN), where two or more core network entities are sharing base stations with non-shared spectrum | SNPN |
| Shared RAN and control plane | NPN is based on 3GPP technology and the RAN is shared with the PLMN. The network control plane is hosted by the PLMN. Data flows remain local in the NPN. | PNI-NPN |
| NPN hosted by the PN | NPN traffic is off-premise but treated differently through Network Slice instances and dedicated data networks. | PNI-NPN |

*Table 4: NPN deployment models*

Different NPN deployment options for TSN integration are investigated, as discussed in-depth in report [5GS21-D54]. TSN integration for the investigated 5G-SMART use cases with an SNPN deployment.

> **KEY RESULTS AND FINDINGS**
> - 5G allows for deployments of non-public networks that are dedicated to an industrial site. Such a deployment may either be independent and standalone for the industrial site. Alternatively it may share functionality with a public mobile network, which may allow for streamlined network deployment and interworking between the public and the non-public networks.

### DEVICE ASPECTS IN THE NETWORK ARCHITECTURE

Given the wide range of 5G industrial applications, diverse communications capabilities would be required for their respective industrial devices. Due to this diversity, communication module and communications technologies used between the different building blocks of a device will have to fulfill the respective demands of the application using the device. It is safe to assume that a single implementation will not fulfill the complete variety of communications characteristics for all applications in an optimized way, considering aspects like power consumption, size, and complexity. On the other hand, implementing special modules for each characteristic profile causes market fragmentation that could prevent the positive effects of economies of scale. Hence, by analyzing the demands of various use cases and respective devices, one can group them into a limited set of device themes based on their communication characteristics. Such industrial 5G devices themes would lay the groundwork for a thorough device classification in the future, and would serve the purpose of:

- Giving guidance to an application designer to choose the suitable device communication capabilities based on the needed traffic patterns and characteristics,
- Identifying the collection of features supported by the respective device,
- Being a foundation for defining test cases within a specific class in order to ensure a specific industrial 5G device class will meet its required E2E performance.

> **KEY RESULTS AND FINDINGS**
> - 5G supports a wide range of different services, from support of time-sensitive communication, support for time synchronization and positioning, etc.. The definition 5G device themes can help application designers to identify devices that are suitable for needs of an industrial application.

### EDGE CLOUD INTEGRATION WITH NON-PUBLIC NETWORKS AND TSN

Edge computing relates to a distributed execution environment (e.g., compute and storage resources) that is close to the location where it is needed in contrast to a remote cloud in some distant data centre. The proximity of the edge premise results in reduced latency between a client and the server application. One typical usage of edge computing in smart manufacturing is to perform tasks on the edge infrastructure that are intensive in complexity, computation, memory and storage. For example, various analytics and monitoring tasks can be executed on the edge, utilizing the cloud capabilities (e.g., scalability, robustness) and process huge amounts of data locally.

It is important to understand how edge computing can be deployed with the different NPN variants of SNPN and PNI-NPN. This has been investigated in [5GS20-D52] and [5GS21-D54] and is summarized in Table 5. In some use cases, the edge computing domain will be deployed locally in the factory premise (referred to as *on-premise standalone edge* in Table 5). Another option is multiple data centres being deployed on-premises (e.g. in separate buildings in the factory), which can handled and managed as a federation (see *on-premise – federated edge* as indicated in Table X ). Such a deployment can offer higher levels of reliability (by means of geo-redundancy) and also increase scalability. A further cloud deployment option (denoted as *on-premise edge integrated with central cloud* in Table 5), where the on-premise edge cloud deployment is jointly handled and managed with a central cloud. Industrial applications can be flexibly located in either the local on-premise deployment or the central deployment, based on service requirements such as latency, reliability and data privacy. All of thses on-premise cloud options are feasible with an SNPN deployment. It is even possible that the NPN itself, which contains a compute infrastructure for its own functionality, can in addition provide the compute platform for the industrial applications, which is referred to as *shared NPN infrastructure* in Table 5.

The different on-premise edge cloud deployments for SNPN are equally feasible for PNI-NPN deployments. In addition, for an PNI-NPN the infrastructure provided by the mobile network operator can serve as edge compute platform for the factory (i.e. *telco edge*) or can provide infrastructure as a service for 3$^{rd}$ party cloud providers.

| NPN deployment options | Possible edge computing deployment |
|---|---|
| Standalone NPN | On-premise – Standalone edge |
| Standalone NPN | On-premise – Federated edge |
| Standalone NPN | On-premise – Integrated edge and central cloud premises |
| Shared NPN infrastructure | On-premise edge |
| PNI-NPN with shared RAN and core control plane | On-premise edge |
| PNI-NPN hosted by the Public network | On-premise edge |
| PNI-NPN hosted by the Public network | Telco/3$^{rd}$ party edge |

*Table 5: Summary of the NPN and edge deployment options*

Using edge computing is also promising to offload time-critical industry applications, like device control (e.g., robot control, AGV control). Besides support for real-time computing in the edge cloud, also reliable low-latency communication is required between the edge cloud and the device, as provided by 5G URLLC and TSN. To provide also deterministic computing within the edge computing platform, the resource management and the workload scheduling must be adjusted to ensure low latency computing guarantees. Since robustness is one crucial factor for industrial use cases, the reliability of such an offloaded application can be improved both in the network and in the edge computing domain. On the network side, IEEE 802.1CB [IEEE17-8021QCB] specifies the Frame Replication and Elimination for Reliability feature of the TSN, which provides the sending of frames on multiple communication paths between a Talker and a Listener by using independent network infrastructure resources. Similarly, in the edge computing domain, the robustness can be increased by deploying multiple instances of the applications on different Pods/nodes. However, there are still open challenges to address on how high-reliability functionality (e.g., TSN FRER) can be realized whent 5G and TSN is integrated with an edge cloud. New concepts for seamless interworking of the TSN FRER functionality with 5G edge computing deployment models have been investigated in report [5GS21-D54].

| KEY RESULTS AND FINDINGS | <ul><li>Edge computing is a promising approach to enable flexibly innovative applications for smart manufacturing.</li><li>In a flexible way, edge computing can be integrated in various 5G non-public network deployments, including local on-premises deployments and deployments with integrated edge and central cloud infrastructure.</li><li>For time-critical industrial applications, real-time computing capabilities are needed by the cloud infrastructure and need to be combined with time-sensitive networking to connect the edge cloud to the physical assets.</li><li>For cloud-based time-critical industrial applications, there it is necessary to integrate the edge cloud with deterministic communication such as TSN, besides the need for real-time computing capabilities of the cloud infrastructure.</li></ul> |
|---|---|

# Industrial-centric network management of the 5G system

Successful integration of a 5G deployment in the smart manufacturing domain needs to consider practical network management for the NPN operator. The 5G system needs to integrate and be jointly operated with other industrial communication network technologies such as Ethernet-based industrial networks including Time-Sensitive Networking. The following points are important to consider when defining a network management platform:

- 5G system network framework should interwork with Ethernet-based industrial network management systems and be usable by industrial applications platforms such as the Robot Operating System (ROS), for which 5GS should provide native connectivity.
- Existing industrial applications and network management platforms shall not require in-depth knowledge of 5G system functionality and mechanism to enable network configuration.
- There is a need for a simplified solution by which existing industrial network technologies can leverage 5G infrastructure functionality (e.g., group management, location, time synchronization) with minimal effort.
- Finally, there should be standardized interfaces that allows ease of configuration between 5G system deployment and existing networking infrastructure.

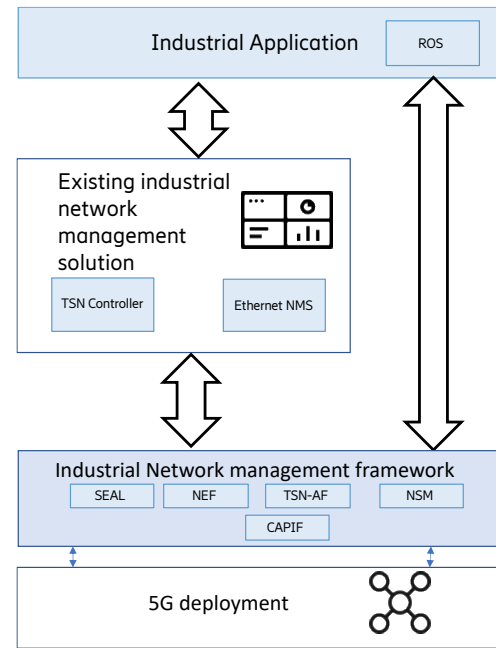

*Figure 20: Industrial-centric network management tool*

To summarize, there is a need for a simplified 5G network management framework that hides the implementation details of the 5G system and provides a simple integration with an existing industrial network management system (NMS) and industrial applications. With this driving mindset, 5G-SMART has proposed the 5G industrial-centric network management framework, as described in report [5GS21-D55] and shown in Figure 20. The network management framework implements an application enablement architecture based on the CAPIF and SEAL components defined by 3GPP. The interaction on the northbound interface is via standardized API defined by 3GPP. Industrial applications such as Robot Operating System (ROS) and existing wired network management framework acts as an API invoker and can implement the API interaction with the framework. The framework is developed based on the assumption that existing OT management applications can implement the consumption of the APIs offered by the proposed network management framework.

| KEY RESULTS AND FINDINGS | ▪ Via the simple APIs provided by the standardized exposure framework defined in SEAL & CAPIF, the 5G system can be easily configured and managed to adapt to the needs of industrial applications and allow integration with existing industrial communication systems. |
|---|---|

# Summary and Outlook

This booklet has outlined key topics to consider in the transformation towards Industry 4.0 taking the results and findings of the 5G-SMART project as its guideline. Important steps towards a successful integration into the ecosystem have been discussed, showing the great potential of introducing 5G but also shedding light into the complex choices when it comes to business relationships, roles, stakeholders, but also deployment options.

The 5G deployments have been described for the three trial sites built up in the 5G-SMART project. The performance evaluations demonstrate the capabilities of 5G in real deployments. A common validation methodology for the trials has been devised which takes a broader angle, moving from 5G capabilities to industry goals, thereby contributing to a wider understand of the quantifiable benefits of 5G in smart manufacturing. Important manufacturing use cases have been discussed, with a special focus on the use cases trialed in the project. The three different testbeds developed in the 5G-SMART project have been used as proof points where the feasibility of realizing smart manufacturing use cases over 5G has been successfully validated, evaluated and demonstrated. Key learnings and findings are highlighted, including challenges encountered.

The results of the investigations of the needed enhancements of 5G and their integration into the ecosystem have been described, where the focus has been on aspects that have a strong business value. In summary, the results and findings outlined in this booklet clearly indicate the potential of 5G in the manufacturing sector. The trials have demonstrated the readiness of 5G for smart manufacturing use cases, but it is also clear that the journey towards Industry 4.0 continues. It is important to move forward by involving the entire ecosystem, where 5G-SMART has made an important contribution.

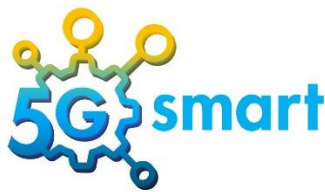
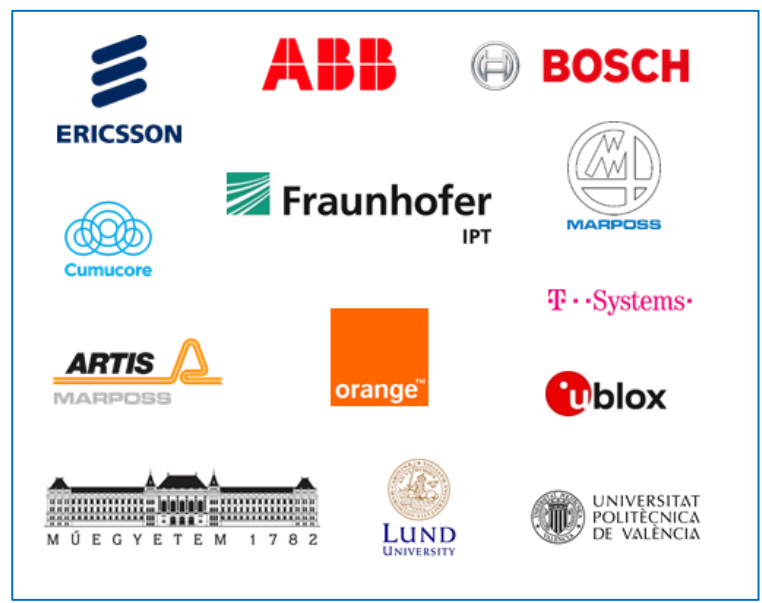

References to all 5G-SMART deliverables, publications as well as presentation and demos can be found at https://5gsmart.eu/.

32# References

| | |
|---|---|
| TR22.804 | Specification # 22.804 (3gpp.org) |
| Lap14 | Laperrière L, Reinhart G. CIRP Encyclopedia of Production Engineering: Springer Berlin; 2014. |
| ISO14a | DIN ISO 22400-1 (2014): Automation systems and integration - Key performance indicators (KPIs) for manufacturing operations management - Part 1: Overview, concepts and terminology. |
| ISO14b | DIN ISO 22400-2 (2014): Automation systems and integration - Key performance indicators (KPIs) for manufacturing operations management - Part 2: Definitions and descriptions. |
| Ans22 | J. Ansari, et al., "Performance of 5G Trials for Industrial Automation" Electronics 11, no. 3: 412, January 2022. https://doi.org/10.3390/electronics11030412 |
| Nor22 | https://www.ericsson.com/en/reports-and-papers/white-papers/5g-spectrum-for-local-industrial-networks#_bookmark9 |
| 5GS20-CT | 5G common terminology, https://5gsmart.eu/wp-content/uploads/5G-SMART-common-terminology.pdf |
| 5GS19-D310 | 5G-SMART, Deliverable 3.1, "Report on industrial shop floor wireless infrastructure", November 2019, https://5gsmart.eu/deliverables/ |
| 5GS20-D110 | 5G-SMART, Deliverable 1.1, "Forward looking smart manufacturing use cases, requirements and KPIs", June 2020, https://5gsmart.eu/deliverables/ |
| 5GS20-D140 | 5G-SMART, Deliverable 1.4, "Radio network deployment options for smart manufacturing", Nov 2020, https://5gsmart.eu/deliverables/ |
| 5GS20-D210 | 5G-SMART, Deliverable 2.1, "Design of 5G-Based Testbed for Industrial Robotics", May 2020, https://5gsmart.eu/deliverables/ |
| 5GS20-D320 | 5G-SMART, Deliverable 3.2, "Report on System Design Options for Monitoring of Workpieces and Machines", May 2020, https://5gsmart.eu/deliverables/ |
| 5GS20-D410 | 5G-SMART, Deliverable D4.1 "Report on design and installation of 5G trial system in Reutlingen", Nov 2020, https://5gsmart.eu/deliverables/ |
| 5GS20-D510 | 5G-SMART, Deliverable 5.1, "Report on new technological features to be supported by 5G standardization and their implementation impact", May 2020, https://5gsmart.eu/deliverables/ |
| 5GS20-D520 | 5G-SMART, Deliverable 5.2, "First report on 5G network architecture options and assessments", Nov 2020, https://5gsmart.eu/deliverables/ |
| 5GS21-D120 | 5G-SMART, Deliverable D1.2, "Analysis of business value creation enabled by 5G for manufacturing industries", May 2021, https://5gsmart.eu/deliverables/ |
| 5GS21-D130 | 5G-SMART, Deliverable D1.3, "Operator business models for smart manufacturing", June 2021, https://5gsmart.eu/deliverables/ |
| 5GS21-D330 | 5G-SMART, Deliverable D3.3, "Report on implementation of options for monitoring of workpiece and machines", May 2021, https://5gsmart.eu/deliverables/ |
| 5GS21-D420 | 5G-SMART, Deliverable D4.2, "Report on 5G Radio Deployability in the Factory", January 2021, https://5gsmart.eu/deliverables/ |
www.5gsmart.eu

| | |
|---|---|
| 5GS21-D430 | 5G-SMART, Deliverable D4.3, "Report on the development of the 5G use cases", May 2021, https://5gsmart.eu/deliverables/ |
| 5GS22-D150 | 5G-SMART, Deliverable D1.5, "Evaluation of radio network deployment options", Dec 2021, https://5gsmart.eu/deliverables/ |
| 5GS22-D220 | 5G-SMART, Deliverable D2.2, "5G-Based Testbed for Trials with Industrial Robotics", January 2022, https://5gsmart.eu/deliverables/ |
| 5GS22-D230 | 5G-SMART, Deliverable D2.3, "Validation of 5G capabilities for industrial robotics", May 2022, https://5gsmart.eu/deliverables/ |
| 5GS22-D340 | 5G-SMART, Deliverable D3.4, "Report on 5G capabilities for enhanced industrial manufacturing processes", May 2022, https://5gsmart.eu/deliverables/ |
| 5GS22-D440 | 5G-SMART, Deliverable D4.4, "Report on validation of the 5G use cases in the factory", May 2022, https://5gsmart.eu/deliverables/ |
| 5GS-LIN | 5G-SMART, LinkedIn page, https://www.linkedin.com/company/5gsmart/ |
| 5GS-SCS | https://5gsmart.eu/standard-contributions/ |
| 5GS-TWI | 5G-SMART, Twitter account, https://twitter.com/5G_smart |
| 5GS-YOU | 5G-SMART, YouTube channel, https://www.youtube.com/channel/UCdhRYuUuSfT97tIivMGLRIg |